\newif\iffullversion{}
\fullversiontrue{}
\fullversionfalse{} 

\iffullversion%
  \documentclass{article}
\else%
  \documentclass[USenglish,oneside,twocolumn]{article}
\fi

\iffullversion%
  \usepackage{amsmath}
\else%
  \usepackage[utf8]{inputenc}
  \usepackage[big]{dgruyter_NEW}
  \usepackage{balance}
  \DOI{foobar}
  \cclogo{\includegraphics{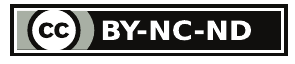}}
\fi%

\usepackage{amsmath}
\usepackage{amssymb,amsthm,amsfonts}
\usepackage{graphicx}
\usepackage{enumitem}
\usepackage{authblk}
\usepackage{listings}
\usepackage{pgfplots}
\pgfplotsset{compat=1.14}
\PassOptionsToPackage{hyphens}{url}\usepackage{hyperref}

\setlist[description]{font=\normalfont\itshape\space} 

\newcommand{\eps}{\varepsilon}
\newcommand{\eeps}{e^\eps}
\newcommand{\emeps}{e^{-\eps}}
\newcommand{\del}{\delta}
\newcommand{\epsdel}{\left(\eps,\del\right)}
\newcommand{\mecha}{\mathcal{M}}
\newcommand{\mech}[1]{\mecha\left(#1\right)}
\newcommand{\numbers}{\mathbb{N}}
\newcommand{\integers}{\mathbb{Z}}
\newcommand{\unitint}{\left[0,1\right]}
\newcommand{\ps}{\pi}
\newcommand{\popt}{\ps_{\textnormal{opt}}}
\newcommand{\mopt}{\mecha_{\textnormal{opt}}}
\newcommand{\pop}[1]{\popt\left(#1\right)}
\newcommand{\pzero}{\ps_{0}}
\newcommand{\pz}[1]{\pzero\left(#1\right)}
\newcommand{\lap}{\textnormal{Lap}}
\newcommand{\plap}{\ps_\lap}
\newcommand{\pgauss}{\ps_{\textnormal{Gauss}}}
\newcommand{\strat}{\rho_\ps}
\newcommand{\slap}{\rho_\lap}
\newcommand{\drop}{\textnormal{drop}}
\newcommand{\keep}{\textnormal{keep}}
\newcommand{\dk}{{\left\{\drop,\keep\right\}}}
\newcommand{\ncr}{n_1}
\newcommand{\nf}{n_2}

\newcommand{\proba}[1]{\mathbb{P}\left[#1\right]}

\newcommand{\expect}[1]{\mathbb{E}\left[#1\right]}

\newcommand{\set}[1]{\left\{#1\right\}}
\newcommand{\setc}[2]{\left\{#1\;\middle|\;#2\right\}}
\newcommand{\card}[1]{\left\lvert#1\right\rvert}

\theoremstyle{plain}
\newtheorem{theorem}{Theorem}
\newtheorem{lemma}{Lemma}
\newtheorem{definition}{Definition}

\lstnewenvironment{sqllisting}{
  \lstset{
    language=SQL,
    morekeywords={select,avg,from,group by,count,sum},
    deletekeywords={value},
    basicstyle=\ttfamily,
    mathescape=true
  }
}{}

\begin{document}

\iffullversion%
  \title{Differentially private partition selection}
  \date{}
\else%
  \title{\huge Differentially private partition selection}
  \runningtitle{Differentially private partition selection}
\fi

\iffullversion%
  \author[1]{Damien Desfontaines}
  \author[2]{James Voss}
  \author[2]{Bryant Gipson}
  \author[2]{Chinmoy Mandayam}

  \affil[1]{Tumult Labs}
  \affil[2]{Google}

  \maketitle
\else%
  \author*[1]{Damien Desfontaines}
  \author[2]{James Voss}
  \author[3]{Bryant Gipson}
  \author[4]{Chinmoy Mandayam}

  \affil[1]{Tumult Labs, damien@desfontain.es}
  \affil[2]{Google, jrvoss@google.com}
  \affil[3]{Google, bryantgipson@google.com}
  \affil[4]{Google, cvm@google.com}
\fi

\begin{abstract}
{Many data analysis operations can be expressed as a GROUP BY query on an
  unbounded set of partitions, followed by a per-partition aggregation. To make
  such a query differentially private, adding noise to each aggregation is not
  enough: we also need to make sure that the set of partitions released is also
  differentially private.
\\
  This problem is not new, and it was recently formally introduced as
  \emph{differentially private set union}~\cite{gopi2020differentially}. In this
  work, we continue this area of study, and focus on the common setting where
  each user is associated with a single partition. In this setting, we propose a
  simple, \emph{optimal} differentially private mechanism that maximizes the
  number of released partitions. We discuss implementation considerations, as
  well as the possible extension of this approach to the setting where each user
  contributes to a fixed, small number of partitions.}
\end{abstract}

\iffullversion%
\else
\maketitle 
\fi%

\section{Introduction}

Suppose that a centralized service collects information on its users, and that
an engineer wants to understand the prevalence of different device models among
the users. They could run a SQL query similar to the following.

\begin{sqllisting}
SELECT
  device_model,
  COUNT(UNIQUE user_id)
FROM database
GROUP BY device_model
\end{sqllisting}

Many common data analysis tasks follow a simple structure, similar to this
example query: a \verb|GROUP BY| operation that defines a set of
\emph{partitions} (here, device models), followed by one or several
aggregations. To make such a query differentially private, it is not enough to
add noise to each count. Indeed, in the example above, suppose that a device
model is particularly rare, and that a single user is associated to this device
model. The presence or absence of this user determines whether this partition
appears in the output: even if the count is noisy, the differential privacy
property is not satisfied. Thus, in addition to the counts, the \emph{set of
partitions present in the output} must also be differentially private. There are
two main ways of ensuring this property.

A first option is to determine the set of output partitions \emph{in advance},
without looking at the private data. In this case, even if some of the
partitions do not appear in the private data, they must still be returned, with
noise added to the zero value. Conversely, if the private data has partitions
that do not appear in the predefined list, they must be dropped from the output.
This option is feasible when grouping users by some fixed categories, or if
partitions can only take a small number of predefined values.

However, this is not always the case. Text-based partitions like search queries
or user agents might take arbitrary values, and often cannot be known without
access to the private dataset. Furthermore, when building a generic DP engine,
usability is paramount, and requiring users to annotate their dataset with all
possible values that can be taken by a given field is a significant burden. This
makes a second option attractive: generating the list of partitions from the
private data itself, in a differentially private way. This problem was formally
introduced in~\cite{gopi2020differentially} as \emph{differentially private set
union}. Each user is associated with one or several partitions, and the goal is
to release as many partitions as possible while making sure that the output is
differentially private.

In~\cite{gopi2020differentially}, the main motivation to study this set union
primitive is natural language processing: the discovery of words and $n$-grams
is essential to these tasks, and can be modeled as a set union problem. In this
context, each user can contribute to many different partitions. In the context
of data analysis queries, however, it is common that each contributes only to a
small number of partitions, often just \emph{one}. This happens when the
partition is a characteristic of each user, for example demographic attributes
or the answer to a survey. In the above SQL query example, if the user ID is a
device identifier, each contributes to at most one device model.

In this work, we focus on this particular single-contribution case, and provide
an \emph{optimal} partition selection strategy for this special case. More
specifically, we show that there is a fundamental upper bound on the probability
of returning a partition associated with $k$ users, and present an algorithm
that achieves this bound.

This paper is structured as follows. After discussing prior work in more detail
and introducing definitions, we present a partition selection mechanism for the
case where each user contributes to one partition, and prove its optimality. We
then discuss possible extensions to cases where each user contributes to
multiple partitions as well as implementation considerations.

\subsection{Prior work}

In this section, we review existing literature on the problem of releasing a set
of partitions from an unbounded set while satisfying differential privacy. This
did not get specific attention until~\cite{gopi2020differentially}, but the
first algorithm that solves it was introduced in~\cite{korolova2009releasing},
for the specific context of privately releasing search queries. This algorithm
works as follows: build a histogram of all partitions, count unique users
associated with each partition, add Laplace noise to each count, and keep only
the partitions whose counts are above a fixed threshold. The scale of the noise
and the value of the threshold determine $\eps$ and $\del$. This method is
simple and natural; it was adapted to work in more general-purpose query engines
in~\cite{wilson2019differentially}. 

In~\cite{gopi2020differentially}, the authors focus on the more general problem
of differentially private set union. The main use case for this work is word and
n-gram discovery in Natural Language Processing: data used in training models
must not leak private information about individuals. In this context, each user
potentially contributes to many elements; the \emph{sensitivity} of the
mechanism can be high. The authors propose two strategies applicable in this
context. First, they use a \emph{weighted} histogram so that if a user
contributes to fewer elements than the maximum sensitivity, these elements can
add more weight to the histogram count. Second, they introduce \emph{policies}
that determine which elements to add to the histogram depending on which
histogram counts are already above the threshold. These strategies obtain
significant utility improvements over the simple Laplace-based strategy.

In this work, in contrast to~\cite{gopi2020differentially}, we focus on the
\emph{low-sensitivity} use case: each user contributes to exactly one element.
This different setting is common in data analysis: when the GROUP BY key
partitions the set of users in distinct partitions, each user can only
contribute to one element to the set union. Choosing the contributions of each
user is therefore not relevant; the only question is to optimize the probability
of releasing each element in the final result. For this specific problem, we
introduce an optimal approach, which maximizes this probability.

\paragraph*{Public partitions}

When the domain of possible partitions is known in advance and considered public
data, no partition selection is necessary. This assumption is typically made
implicitly in existing work on histogram publication, either by assuming that
the domain is known exactly and not too
large~\cite{hay2009boosting,ding2011differentially,xiao2010differential,xiao2012dpcube,acs2012differentially,xu2013differentially,zhang2017privbayes},
or that the attributes are numeric and
bounded~\cite{li2014differentially}. In the former case, no partition selection
is necessary; the strategy usually revolves around grouping known partitions
together to limit the impact of the noise. In the latter case, the possible
partitions are also indirectly known in advance (all possible intervals in a
fixed numerical range), and the problem is to find which intervals to use to
slice the data. With such pre-existing knowledge about the partitions, our
approach does not provide any benefit.

\paragraph*{Domain of fixed size}

When the domain does not conform to one of the assumptions described above, the
data domain might still be a subset of some large domain. For example, integer
attributes are typically stored using 64 bits. Similarly, it is reasonable to
assume that search queries or URLs are strings whose size is bounded by some
large number.

We can use this fact to perform partition selection by adding noise to all
possible partitions, including the ones that do not contain any data, and only
return the ones that are above a given threshold. This process can be simulated
in an efficient way, without actually enumerating all
partitions~\cite{cormode2011differentially}. Other methods might be possible;
for example, one could imagine simulating the sparse vector
technique~\cite{dwork2009complexity} or one of its multiple-queries
variants~\cite{dwork2014algorithmic,lee2014top,lyu2016understanding,kaplan2020sparse}
to ask the number of users in all possible partitions, while ignoring the
privacy cost of answers below a threshold.

We are not aware of any work using these techniques for the specific problem of
partition selection. We also postulate that they are likely to fail for
extremely large domain sizes (like long strings); the technique
in~\cite{cormode2011differentially} outputs a number of false positive
partitions linear in the domain size.

\paragraph*{Differences with our approach}

In this work, we focus on cases where all assumptions above fail because the
domain of the data is unbounded or too large. As such, the only way to learn
this domain is by looking at the private data, which must be done in a
differentially private way. This assumption is particularly suited to building
generic tooling, like general-purpose differentially private query
engines~\cite{bater2018shrinkwrap,johnson2018towards,wilson2019differentially,kotsogiannis2019privatesql}.
Indeed, to use such an engine, either all the domain of the input data must be
enumerated in advance, or partition selection is necessary. But this requires
the analyst or data owner to document the data domain for all input databases.
This is a significant usability burden, which makes it difficult to scale the
use of the query engine. This problem is the main motivator for our work.

\subsection{Definitions}

Differential privacy (DP) is a standard notion to quantify the privacy
guarantees of statistical data. For the problem of differentially private set
union, we use $\epsdel$-DP\@.

\begin{definition}[Differential privacy~\cite{dwork2006calibrating}]
  A randomized mechanism $\mecha$ is $\epsdel$-differentially private if for any
  two databases $D$ and $D'$, where $D'$ can be obtained from $D$ by either
  adding or removing one user, and for all sets $S$ of possible outputs:
  \begin{align*}
    \proba{\mech{D}\in S} \le e^\eps \proba{\mech{D'}\in S} + \del.
  \end{align*}
\end{definition}

Let us formalize the problem addressed in this work.

\begin{definition}[Differentially private partition selection]\label{def:dpps}
  Let $U$ be a universe of \emph{partitions}, possibly infinite. A
  \emph{partition selection mechanism} is a mechanism $\mecha$ that takes a
  database $D$ in which each user $i$ contributes a subset $W_i\subset U$ of
  partitions, and outputs a subset $\mech{D}\subseteq\cup_i W_i$.

  The problem of \emph{differentially private partition selection}\footnote{Also
  called \emph{differentially private set union}~\cite{gopi2020differentially}.}
  consists in finding a mechanism $\mecha$ that outputs as many partitions as
  possible while satisfying $\epsdel$-differential privacy.
\end{definition}

In the main section of this paper, we assume that each user contributes to only
one partition ($|W_i|=1$ for all $i$). We first study the simplified problem of
considering each partition independently. The only question then is: with which
probability do we release this partition? And the strategy can simply be reduced
to a function associating the number of users in a partition with the
probability of keeping the partition. After finding an optimal primitive for
this simpler problem, we show that it is actually optimal in a stronger sense,
even among mechanisms that consider all partitions simultaneously.

\begin{definition}[Partition selection primitive]
  A partition selection primitive is a function
  $\pi:\numbers\rightarrow\unitint$ such that $\ps(0)=0$. The corresponding
  partition selection strategy consists in counting the number $n$ of users in
  each partition, and releasing this partition with probability $\ps(n)$.

  We say that a partition selection primitive is $\epsdel$-differentially
  private if the corresponding partition selection strategy
  $\strat:\numbers\rightarrow\dk$, defined by:
  \[
  \strat(n) =
    \begin{cases}
      \drop & \textnormal{with probability } 1-\ps(n) \\
      \keep & \textnormal{with probability } \ps(n)
    \end{cases}
  \]
  is $\epsdel$-differentially private.
\end{definition}

Note that partitions associated with no users are not present in the input data,
so the probability of releasing them has to be $0$: the definition requires
$\ps(0)=0$.

\section{Main result}

In this section, we define an $\epsdel$-DP partition selection primitive
$\popt$ and prove that the corresponding partition selection strategy is
optimal. In this context, optimal means that it maximizes the probability of
releasing a partition with $n$ users, for all $n$.

\begin{definition}[Optimal partition selection primitive]\label{def:optimal}
  A partition selection primitive $\popt$ is \emph{optimal for $\epsdel$-DP} if
  it is $\epsdel$-DP, and if for all $\epsdel$-DP partition selection primitives
  $\ps$ and all $n\in\numbers$:
  \[
    \ps(n) \le \pop{n}.
  \]
\end{definition}

We introduce our main result, then we prove it in two steps: we first prove that
the optimal partition selection primitive can be obtained recursively, then
derive the closed-form formula of our main result from the recurrence relation.

\begin{theorem}[General solution for $\popt$]\label{thm:general-solution}
  Let $\eps>0$ and $\del\in(0,1)$. Defining:
  \begin{align*}
    \ncr & = 1+\left\lfloor\frac{1}{\eps}\ln\left(
                \frac{\eeps+2\del-1}{(\eeps+1)\del}\right)
             \right\rfloor, \\
    \nf  & = \ncr + \left\lfloor\frac{1}{\eps}\ln\left(
                1+\frac{\eeps-1}{\del}\left(1-\pop{\ncr}\right)
             \right)\right\rfloor,
  \end{align*}
  and $m=n-\ncr$, the partition selection primitive $\popt$ defined by:
  \iffullversion%
    \[
      \pop{n} =
        \begin{cases}
          \frac{e^{n\eps}-1}{\eeps-1}\cdot\del 
            & \textnormal{if } n \le \ncr \\
          \left(1-e^{-m\eps}\right)\left(1+\frac{\del}{\eeps-1}\right)+e^{-m\eps}\pop{\ncr}
            & \textnormal{if } n > \ncr \textnormal{\ and } n \le \nf \\
          1 & \textnormal{otherwise}
        \end{cases}
    \]
  \else%
    \begin{itemize}
      \item $\pop{n}=\frac{e^{n\eps}-1}{\eeps-1}\cdot\del $ if $n\le\ncr$,
      \item $\pop{n}=\left(1-e^{-m\eps}\right)\left(1+\frac{\del}{\eeps-1}\right)+e^{-m\eps}\pop{\ncr}$
        if $n>\ncr$ and $n\le\nf$,
      \item $1$ otherwise
    \end{itemize}
  \fi%
  is optimal for $\epsdel$-DP\@.
\end{theorem}

These formulas assume $\eps>0$ and $\del>0$. We also cover the special cases
where $\eps=0$ or $\del=0$.

\begin{theorem}[Special cases for $\popt$]\label{thm:special-cases}
  \!
  \begin{enumerate}
    \item If $\del=0$, partition selection is impossible: the optimal partition
      selection primitive $\popt$ for $\left(\eps,0\right)$-DP is defined by
      $\pop{n}=0$ for all $n$.
    \item If $\eps=0$, the optimal partition selection primitive $\popt$ for
      $\left(0,\del\right)$-DP is defined by $\pop{n}=\min\left(1,n\del\right)$
      for all $n$.
  \end{enumerate}
\end{theorem}

\subsection{Recursive construction}\label{sec:recursive}

How do we construct a partition selection primitive $\pi$ so that the partition
is output with the highest possible probability under the constraint that $\pi$
is $\epsdel$-DP\@? Using the definition of differential privacy, the following
inequalities must hold for all $n\in\numbers$.

\begin{align}
  \pi(n+1) &\leq \eeps \pi(n) + \del \label{eq:pn1-lhs-ineq} \\
  \pi(n) &\leq \eeps \pi(n+1) + \del \label{eq:pn-lhs-ineq} \\
  (1-\pi(n+1)) &\leq \eeps (1-\pi(n)) + \del \label{eq:pn1c-lhs-ineq} \\
  (1-\pi(n)) &\leq \eeps (1-\pi(n+1)) + \del. \label{eq:pnc-lhs-ineq}
\end{align}

These inequalities are not only necessary, but also \emph{sufficient} for $\pi$
to be DP\@. Thus, the optimal partition selection primitive can be constructed
by recurrence, maximizing each value while still satisfying the inequalities
above. As we will show, only inequalities~\eqref{eq:pn1-lhs-ineq}
and~\eqref{eq:pnc-lhs-ineq} above need be included in the recurrence
relationship. The latter can be rearranged as:
\[
  \pop{n+1} \leq 1 - \emeps(1-\pop{n}-\del)
\]
which leads to the following recursive formulation for $\popt$.

\begin{lemma}[Recursive solution for $\popt$]\label{lem:recursive-solution}
  Given $\del \in \unitint$ and $\eps \geq 0$, $\popt$ satisfies the following
  recurrence relationship: $\pop{0}=0$, and for all $n\ge0$:
  \iffullversion%
    \begin{equation}\label{eq:pi_opt-recursive-defn}
      \pop{n+1} =
        \min\left(
          \eeps \pop{n} + \del,
          1 - \emeps(1- \pop{n} - \del),
          1
        \right)
    \end{equation}
  \else%
    \begin{equation}\label{eq:pi_opt-recursive-defn}
      \begin{aligned}
        \pop{n+1} =
          & \min(\\
          & \quad \eeps\pop{n} + \del, \\
          & \quad 1-\emeps(1-\pop{n}-\del), \\
          & \quad 1)
      \end{aligned}
    \end{equation}
  \fi%
  \begin{proof}
    Let $\pzero$ be defined by recurrence as above; we will prove that
    $\pzero=\popt$.

    First, let us show that $\pzero$ is monotonic. Fix $n\in\numbers$. It
    suffices to show for each argument of the min function
    in~\eqref{eq:pi_opt-recursive-defn} is larger than $\pz{n}$.
    \begin{description}
      \item [First argument.] Since $\eps\ge0$ implies $\eeps\ge1$ and
        $\del\ge0$, we trivially have $\eeps\pz{n}+\del\ge\pz{n}$.
      \item [Second argument.] We have:
        \begin{align*}
          1 - \emeps(1 - \pz{n} -\del)
          & =   1 - \emeps(1-\pz{n}) + \emeps\del \\
          & \ge 1 - (1-\pz{n}) \\
          & = \pz{n}
        \end{align*}
        using that $1-\pz{n}\ge0$ since $\pz{n}\le1$
        by~\eqref{eq:pi_opt-recursive-defn}.
      \item [Third argument.] This is immediate
        given~\eqref{eq:pi_opt-recursive-defn} and the fact that $\pz{0}=0$.
    \end{description}
    It follows that $\pz{n+1}\ge\pz{n}$.

    Because $\pzero$ is monotonic, it immediately satisfies
    inequalities~\eqref{eq:pn-lhs-ineq} and~\eqref{eq:pn1c-lhs-ineq}, and
    inequalities~\eqref{eq:pn1-lhs-ineq} and~\eqref{eq:pnc-lhs-ineq} are
    satisfied by definition.

    Since $\pzero$ satisfies all four inequalities above, it is $\epsdel$-DP\@.
    Its optimality follows immediately by recurrence: for each $n+1$, if
    $\ps(n+1)>\pop{n+1}$, it cannot be $\epsdel$-DP, as one of the inequalities
    above is not satisfied: $\pzero$ is the fastest-growing DP partition
    selection strategy, and therefore equal to $\popt$.
  \end{proof}
\end{lemma}

Note that the special cases for $\popt$ in Theorem~\ref{thm:special-cases} can
be immediately derived from Lemma~\ref{lem:recursive-solution}.

\subsection{Derivation of the closed-form solution}

Let us now show that the closed-form solution of
Theorem~\ref{thm:general-solution} can be derived from the recursive solution
in~\ref{lem:recursive-solution}. First, we show that there is a crossover point
$\ncr$, below which only the first term of the recurrence relation matters, and
after which only the second term matters (until $\pop{n}$ reaches $1$).

\begin{lemma}\label{lem:crossover-points}
  Assume $\eps>0$ and $\del>0$. There are crossover points $\ncr,\nf\in\numbers$
  such that $0<\ncr\le\nf$ and
  \iffullversion%
    \begin{equation}
      \pop{n} =
        \begin{cases}
          0 & \textnormal{if } n=0 \\
          \pop{n-1}\eeps + \del
            & \textnormal{if } n > 0 \textnormal{\ and } n \le \ncr \\
          1-\emeps\left(1-\pop{n-1}-\del\right)
            & \textnormal{if } n > \ncr \textnormal{\ and } n \le \nf \\
          1 & \textnormal{otherwise}
        \end{cases}
    \end{equation}
  \else%
    \begin{itemize}
      \item $\pop{n}=0$ if $n=0$,
      \item $\pop{n}=\pop{n-1}\eeps+\del$ if $n>0$ and $n\le\ncr$,
      \item $\pop{n}=1-\emeps\left(1-\pop{n-1}-\del\right)$ if $n>\ncr$ and
        $n\le\nf$,
      \item $\pop{n}=1$ otherwise.
    \end{itemize}
  \fi%

  \begin{proof}
    We consider the arguments in the min statement
    in~\eqref{eq:pi_opt-recursive-defn}, substituting $x$ for $\pop{n}$:
    \begin{align*}
      \alpha_1(x) & = \eeps x + \del \\
      \alpha_2(x) & = 1 - \emeps(1 - x - \del) \\
      \alpha_3(x) & = 1
    \end{align*}
    This substitution allows us to work directly in the space of probabilities
    instead of restricting ourselves to the sequence
    ${\left(\pop{n}\right)}_{n=0}^{\infty}$. Taking the first derivative of
    these functions yields:
    \begin{align*}
      \alpha_1^\prime(x) &= \eeps \\
      \alpha_2^\prime(x) &= \emeps \\
      \alpha_3^\prime(x) &= 0
    \end{align*}
    Since the derivative of $\alpha_1(x)-\alpha_2(x)$ is $\eeps-\emeps>0$, there
    exists at most one crossover point $x_1$ such that $\alpha_1(x)<\alpha_2(x)$
    for all $x<x_1$, $\alpha_2(x_1)=\alpha_1(x_1)$, and
    $\alpha_1(x)>\alpha_2(x)$ for all $x>x_1$. Setting
    $\alpha_1(x)=\alpha_2(x)$ and solving for $x$ yields:
    \[
      \eeps x + \del = 1 - \emeps(1-x-\del)
    \]
    which leads to:
    \[
       \eeps x - \emeps x = 1 - \del - \emeps(1-x-\del)
    \]
    and finally:
    \[
      x_1=(1-\del)\cdot\frac{1-\emeps}{\eeps-\emeps}.
    \]
    Since the derivative of $\alpha_2(x)-\alpha_3(x)$ is $\emeps>0$, there
    exists at most one crossover point $x_2$ such that $\alpha_2(x)<\alpha_3(x)$
    for all $x<x_2$, $\alpha_2(x_2)=\alpha_3(x_2)$, and
    $\alpha_2(x)>\alpha_3(x)$ for all $x>x_2$. Setting $\alpha_2(x)=\alpha_3(x)$
    and solving for $x$ yields: \[
      x_2 = 1-\del.
    \]
    From the formulas for $x_1$ and $x_2$, it is immediate that $0<x_1<x_2<1$.
    As such, the interval $\unitint$ can be divided into three non-empty
    intervals:
    \begin{enumerate}
      \item On $\left[0, x_1\right]$, $\alpha_1(x)$ is the active argument of $\min(\alpha_1(x),\alpha_2(x),\alpha_3(x))$.
      \item On $\left[x_1, x_2\right]$, $\alpha_2(x)$ is the active argument of $\min(\alpha_1(x),\alpha_2(x),\alpha_3(x))$.
      \item On $\left[x_2, 1\right]$, $\alpha_3(x)$ is the active argument of $\min(\alpha_1(x),\alpha_2(x),\alpha_3(x))$.
    \end{enumerate}

    The existence of the crossover points is not enough to prove the lemma: we
    must also show that these points are reached in a finite number of steps.
    For all $n \ge 1$ such that $\pop{n}\neq1$, we have:
    \iffullversion%
      \begin{align*}
        & \pop{n}-\pop{n-1} \\
        & \qquad =   \min\left(\eeps\pop{n-1}+\del, 1-\emeps\left(1-\pop{n-1}-\del\right)\right)-\pop{n-1} \\
        & \qquad \ge \min\left(\del,\left(1-\emeps\right)\left(1-\pop{n-1}\right)+\emeps\del\right) \\
        & \qquad \ge \emeps\del.
      \end{align*}
    \else%
      \begin{align*}
        & \pop{n}-\pop{n-1} \\
        & \qquad = \min( \\
        & \qquad \qquad \eeps\pop{n-1}+\del, \\
        & \qquad \qquad 1-\emeps\left(1-\pop{n-1}-\del\right) \\
        & \qquad \vphantom{=} )-\pop{n-1} \\
        & \qquad \ge \min\left(\del,\left(1-\emeps\right)\left(1-\pop{n-1}\right)+\emeps\del\right) \\
        & \qquad \ge \emeps\del.
      \end{align*}
    \fi%
    Since $\pop{n}-\pop{n-1}$ is bounded from below by a strictly positive
    constant $\emeps\del$, the sequence achieves the maximal probability 1 for
    finite $n$.
  \end{proof}
\end{lemma}

This allows us to derive the closed-form solution for $n<\ncr$ and for
$\ncr \le n < \nf$ stated in Theorem~\ref{thm:general-solution}.

\begin{lemma}\label{lem:before-ncr}
  Assume $\eps>0$ and $\del\le0$. If $n\leq\ncr$, then
  $\pop{n}=\frac{e^{n\eps}-1}{\eeps-1}\cdot\del$. If $\ncr \le n < \nf$, then
  denoting $m=n-\ncr$:
  \[
    \pop{n}=\left(1-e^{-m\eps}\right)\left(1+\frac{\del}{\eeps-1}\right)+e^{-m\eps}\pop{\ncr}.
  \]
  \begin{proof}
    For $n<\ncr$, expanding the recurrence relation yields:
    \begin{align*}
      \pop{n}
        & = \pop{n-1}\eeps+\del \\
        & = \del\sum_{k=0}^{n-1}e^{k\eps} \\
        & = \frac{e^{n\eps}-1}{\eeps-1}\cdot\del.
    \end{align*}
    For $\ncr \le n < \nf$, denoting $m=n-\ncr$, expanding the recurrence
    relation yields:
    \begin{align*}
      \pop{n}
        & = 1-\emeps\left(1-\pop{n-1}-\del\right) \\
        & = \left(1-\emeps+\del\emeps\right)\sum_{k=0}^{m-1}e^{-k\eps}+e^{-m\eps}\pop{\ncr} \\
        & = \left(1-\emeps+\del\emeps\right)\frac{1-e^{-m\eps}}{1-\emeps}+e^{-m\eps}\pop{\ncr} \\
        & = \left(1-e^{-m\eps}\right)\left(1+\frac{\del}{\eeps-1}\right)+e^{-m\eps}\pop{\ncr}.
    \end{align*}
  \end{proof}
\end{lemma}

We can now find a closed-form solution for $\ncr$ and for $\nf$.

\begin{lemma}\label{lem:ncr}
  The first crossover point $\ncr$ is:
  \begin{equation}
  \ncr = 1+\left\lfloor \frac{1}{\eps} \ln\left(
    \frac{\eeps+2\del-1}{\del(\eeps+1)} \right)\right\rfloor \label{eq:n1}
  \end{equation}
  \begin{proof}
    Using the formula for $x_1$ in the proof of
    Lemma~\ref{lem:crossover-points}, we see that $\pop{n-1}\le x_1$ whenever:
    \[
      \frac{e^{(n-1)\eps}-1}{\eeps-1}\cdot\del\le\frac{1-\del}{\eeps+1}.
    \]
    Rearranging terms, we can rewrite this inequality as:
    \begin{align*}
      n & \le 1 + \frac{1}{\eps}\ln\left[\frac{(1-\del)(\eeps-1)}{\del(\eeps+1)}+1\right] \\
        & =   1 + \frac{1}{\eps}\ln\left[\frac{(1-\del)(\eeps-1)+\del(\eeps+1)}{\del(\eeps+1)}\right] \\
        & =   1 + \frac{1}{\eps}\ln\left[\frac{\eeps+2\del-1}{\del(\eeps+1)}\right].
    \end{align*}
    Since $n$ is an integer, the supremum value defining $\ncr$ is achieved by
    taking the floor of the right-hand side of this inequality, which concludes
    the proof.
  \end{proof}
\end{lemma}

\begin{lemma}\label{lem:nf}
  The second crossover point $\nf$ is:
  \[
    \nf = \ncr + \left\lfloor\frac{1}{\eps}\ln\left(
                1+\frac{\eeps-1}{\del}\left(1-\pop{\ncr}\right)
          \right)\right\rfloor
  \]
  \begin{proof}
    We want to find the maximal $m$ such that:
    \[
      \left(1-e^{-m\eps}\right)\left(1+\frac{\del}{\eeps-1}\right)+e^{-m\eps}\pop{\ncr} \le 1.
    \]
    We can rewrite this condition into:
    \[
      -e^{-m\eps}\left(1+\frac{\del}{\eeps-1}-\pop{\ncr}\right) \le \frac{-\del}{\eeps-1}
    \]
    which leads to:
    \begin{align*}
      e^{m\eps}
        & \le \frac{\eeps-1}{\del}\left(1+\frac{\del}{\eeps-1}-\pop{\ncr}\right) \\
        & \le 1+\frac{\eeps-1}{\del}\left(1-\pop{\ncr}\right)
    \end{align*}
    and finally:
    \[
      m \le \frac{1}{\eps}\ln\left(1+\frac{\eeps-1}{\del}\left(1-\pop{\ncr}\right)\right)
    \]
    since $m$ must be an integer, we take the floor of the right-hand side of
    this inequality to obtain the result.
  \end{proof}
\end{lemma}

\subsection{More generic optimality result}

Theorem~\ref{thm:general-solution} provides an optimal partition selection
primitive in the sense of Definition~\ref{def:optimal}: a mechanism using this
primitive on each partition separately is optimal among the class of mechanisms
that consider every partition separately. The mechanism cannot use auxiliary
knowledge about relationships within partitions, and the decision for a given
partition cannot depend on the data in other partitions. Can we extend the
optimality result to a larger class of algorithms, that take the full list of
partitions as input?

We can answer that question in the affirmative, in the particular case where
each user contributions a single partition. First, we need to define what
optimality means in a more general context. Recall that a partition selection
mechanism takes a database $D$ in which each user contributes a subset
$W_i\subset U$ of partitions, and outputs a subset $\mech{D}\subseteq\cup_i
W_i$. The goal is to output as many partitions as possible, which we capture by
maximizing the expected value of the number of output partitions.

\begin{definition}[Optimal partition selection mechanism]
  A partition selection mechanism $\mecha$ is \emph{optimal for $\epsdel$-DP and
  sensitivity $\kappa$} if it is $\epsdel$-DP, and if for all $\epsdel$-DP
  partition selection mechanisms $\mecha'$, and all databases $D$ in which each
  user contributes at most $\kappa$ partitions:
  \[
    \expect{\card{\mecha'\left(D\right)}} \le \expect{\card{\mech{D}}}.
  \]
\end{definition}

We can now prove our more generic optimality result.

\begin{theorem}
  Let $\mopt$ be the partition selection mechanism that, on input $D$, returns
  each partition $k$ with probability $\pop{\card{\setc{i}{W_i=\set{k}}}}$. Then
  $\mopt$ is optimal for $\epsdel$-DP and sensitivity $1$.
  \begin{proof}
    Let $\mecha$ be a partition selection mechanism. Since we assume that every
    user contributes to at most one partition ($\kappa=1$), it is equivalent to
    consider the input of $\mecha$ to be the histogram
    $\left(n_i\right)_{i\in U}$, where $n_i$ is the number of users associated
    to partition $i$. Of course, if $n_k=0$ for some $k$, then $k$ must not be
    in the output set.

    Now, for a given partition $k$, fix all values of the histogram except
    $n_k=n$, and denote $f(n)=\proba{k\in\mech{\left(n_i\right)_{i\in U}}}$.
    Then $f(n)$ must satisfy inequalities~\ref{eq:pn1-lhs-ineq}
    to~\ref{eq:pnc-lhs-ineq} from Section~\ref{sec:recursive} in order for
    $\mecha$ to be $\epsdel$-DP. Then, by Theorem~\ref{thm:general-solution},
    $f(n)\le\popt(n)$. Now, for a given input $\left(n_i\right)_{i\in U}$, the
    expected size of the output set is given by:
    \[
      \sum_{k\in U} \proba{k\in\mech{\left(n_i\right)_{i\in U}}}
    \]
    which is bounded by $\sum_{k\in U} \pop{n_k}$. This is exactly the expected
    output size obtained with $\mopt$, which concludes the proof. 
  \end{proof}
\end{theorem}

\section{Thresholding interpretation}

In this section, we show that modulo a minor change in $\eps$ or $\del$, the
optimal partition selection primitive $\popt$ can be interpreted as a
\emph{noisy thresholding} operation, similar to the Laplace-based strategy, but
using a truncated version of the geometric distribution. We first define this
distribution, then we use it to prove this second characterization of $\popt$.

\iffullversion%
  \newcommand{\tsgd}{Truncated symmetric geometric distribution}
\else%
  \newcommand{\tsgd}{$k$-TSGD}
\fi

\begin{definition}[\tsgd]
  Given $p\in(0,1)$ and $k \in \numbers$ such that $k \ge1$, the \emph{$k$-truncated
  symmetric geometric distribution} ($k$-TSGD) of parameter $p$
  is the distribution defined on $\integers$ such that:
  \begin{equation}
    \proba{X=x} =
      \begin{cases}
        c \cdot {(1-p)}^{|x|} & \textnormal{if } x\in\left[-k,k\right]\cap \integers\\
        0                            & \textnormal{otherwise}
      \end{cases} \label{eq:truncated-geometric-distribution}
  \end{equation}
  where $c = \frac{p}{1 + (1-p) -2{(1-p)}^{k+1}}$ is a normalization constant
  ensuring that the total probability is $1$.
\end{definition}

This distribution can also be obtained by taking a symmetric two-sided geometric
distribution\footnote{Also called the discrete Laplace
distribution~\cite{inusah2006discrete}.}~\cite{ghosh2012universally}, with
success probability $p$ and conditioning on the event that the result is in
$\left[-k,k\right]$. As such, the $k$-truncated symmetric geometric distribution
is the discrete analogue of the truncated Laplace
distribution~\cite{geng2020tight}. A similar construction was also defined
in~\cite{geng2016adp} to prove a lower bound on loss with $\epsdel$-differential
privacy, but is not a proper probability distribution, since its total mass does
not sum up to one\footnote{In the proof of Theorem~8, the sum for non-negative
$i$ is assumed to sum up to $1/2$, but $0$ is counted twice when summing
non-negative and non-positive $i$.}.

Given privacy parameters $\eps$ and $\del$, we can set the values of $p$ and $k$
such that adding noise drawn from the truncated geometric distribution
achieves $\epsdel$-differential privacy for counting queries.

\begin{definition}[Truncated geometric mechanism]\label{def:tgm}
  Given privacy parameters $\eps>0$ and $\del>0$, let $p = 1-\emeps$ and
  $k = \left\lceil \frac{1}{\eps}\ln\left(\frac{\eeps+2\del-1}{(\eeps+1)\del}
  \right) \right\rceil$. Let the true result of an integer-valued query with
  sensitivity 1 be $\mu \in \integers$. Then the truncated geometric mechanism
  returns $\mu + X$, where $X$ is drawn from the $k$-TSGD with success
  probability $p$.  The result has the distribution:
  \[
    \proba{Y=y} =
      \begin{cases}
        c \cdot e^{-|y-\mu|\eps} & \textnormal{if } y\in\left[\mu-k,\mu+k\right]\cap \integers \\
        0                       & \textnormal{otherwise}
      \end{cases}
  \]
  where $c = \frac{1-\emeps}{1 + \emeps -2e^{-(k+1)\eps}}$ is a normalization
  constant ensuring that the total probability is $1$.
\end{definition}

The value of $k$ is the smallest value such that
\[
  \proba{X=k}=\frac{e^{-k\eps}(1 - \emeps)}{1 + \emeps - 2e^{-(k+1)\eps}}\le\del
\]
for the $k$-TSGD\@.

\begin{theorem}\label{thm:tgm-is-dp}
  The truncated geometric mechanism satisfies $\epsdel$-differential privacy.
  \begin{proof}
    This follows the same line of reasoning as the proof of Theorem~1
    in~\cite{geng2020tight}. The only difference is the change from a continuous
    distribution to a discrete distribution, since all the values are integers.
    If the result of the query before adding noise is $\mu$, then for an
    adjacent database, the corresponding value $\mu'$ must be in
    $\set{\mu-1,\mu,\mu+1}$. If $\mu'=\mu$, the distribution of the output after
    adding noise is unchanged, trivially satisfying the $\epsdel$-differential
    privacy property. By symmetry, it is sufficient to analyze the case when
    $\mu'=\mu+1$. Here, the new distribution of the output of the mechanism is
    \iffullversion%
      \[
        \proba{Y'=y} =
          \begin{cases}
            c \cdot e^{-|y-\mu-1|\eps} & \textnormal{if } y\in\left[\mu-k + 1,\mu+k +1\right]\cap \integers \\
            0                       & \textnormal{otherwise}
          \end{cases}
      \]
    \else%
      \[
        \proba{Y'=y} = c \cdot e^{-|y-\mu-1|\eps}
      \]
      if $y\in\left[\mu-k + 1,\mu+k +1\right]\cap\integers$, and $\proba{Y'=y}$
      otherwise.
    \fi%

    By symmetry, to show that $\epsdel$-differential privacy is satisfied, we
    only need to show that $\proba{Y'\in S} \le \eeps\proba{Y\in S} +\delta$ for
    all $S\subset\integers$. For all values $y \in \integers$ except $\mu+k+1$,
    $\proba{Y'=y}\le\eeps\proba{Y=y}$. Also, $\proba{Y=\mu+k+1}=0$ and
    $\proba{Y'=\mu+k+1}=\proba{X=k}>0$. Therefore,
    $\proba{Y'\in S}-\eeps\proba{Y\in S}$ is maximized when $S=\set{\mu+k+1}$.
    This means that the condition is satisfied if $\proba{X=k} \le \delta$. From
    the definition of $k$ in the truncated geometric mechanism:
    \iffullversion%
      \begin{alignat*}{3}
        & & k &= \left\lceil \frac{1}{\eps}\ln\left(\frac{\eeps+2\del-1}{(\eeps+1)\del}
        \right) \right\rceil\\
        &\implies & e^{k\eps} &\ge \frac{\eeps+2\del-1}{(\eeps+1)\del}\\
        &\implies &(e^{k\eps})(\eeps+1)\del - 2\del &\ge \eeps-1\\
        &\implies &(1 + \emeps - 2e^{-(k+1)\eps})\del &\ge e^{-k\eps}(1 - \emeps)\\
        &\implies &\del &\ge
           \frac{e^{-k\eps}(1 - \emeps)}{1 + \emeps - 2e^{-(k+1)\eps}}\\
        &\implies &\del &\ge \proba{X=k}.
      \end{alignat*}
    \else%
      \[
        k=\left\lceil\frac{1}{\eps}\ln\left(\frac{\eeps+2\del-1}{(\eeps+1)\del}\right)\right\rceil
      \]
      which leads to:
      \[
        e^{k\eps} \ge \frac{\eeps+2\del-1}{(\eeps+1)\del}
      \]
      thus:
      \[
        (e^{k\eps})(\eeps+1)\del - 2\del \ge \eeps-1
      \]
      and:
      \[
        (1 + \emeps - 2e^{-(k+1)\eps})\del \ge e^{-k\eps}(1 - \emeps)
      \]
      and finally:
      \begin{align*}
        \del & \ge \frac{e^{-k\eps}(1 - \emeps)}{1 + \emeps - 2e^{-(k+1)\eps}} \\
             & \ge \proba{X=k}.
      \end{align*}
    \fi%
  \end{proof}
\end{theorem}

Let us get some intuition why thresholding the truncated geometric mechanism
leads to an optimal partition selection primitive. First, we compute the tail
cumulative distribution function for the output of the truncated geometric
mechanism. Summing the probability masses gives a geometric series:
\begin{equation*}
  \proba{Y \geq y} =
    \begin{cases}
      1 &\textnormal{if } y \leq \mu-k \\
      1 - \frac {e^{(k+y-\mu)\epsilon} -1}{e^\epsilon -1} ce^{-k\epsilon} &\textnormal{if } \mu-k \leq y \leq \mu -1\\
      \frac {e^{(\mu+k+1 -y)\epsilon} -1}{e^\epsilon -1} ce^{-k\epsilon} &\textnormal{if } \mu \leq y \leq \mu+k \\
      0 &\textnormal{if } y > \mu+k.
    \end{cases}
\end{equation*}

If we define $\delta = ce^{-k\epsilon}$ and rearrange the the cases as functions
of $\mu$, we get:

\begin{equation}
\proba{Y \geq y} =
\begin{cases}
0 &\textnormal{if } \mu < y -k \\
\frac {e^{(\mu+k+1 -y)\epsilon} -1}{e^\epsilon -1} \del &\textnormal{if }  y-k \leq \mu \leq y \\
1 - \frac {e^{(k+y-\mu)\epsilon} -1}{e^\epsilon -1} \del &\textnormal{if } y+1 \leq \mu \leq y+k \\
1 &\textnormal{if } \mu \geq y+k. \label{eq:thresholding-prob}
\end{cases}
\end{equation}

The $\mu \leq y$ cases of the formula are the same as the closed-form formula for
$\popt$ in Theorem~\ref{thm:general-solution} for values less than $\ncr$. The
$\mu > y$ cases are simply the symmetric reflection of the former. We formalize
this intuition and show that whenever
$\frac{1}{\eps}\ln\left(\frac{\eeps+2\del-1}{(\eeps+1)\del}\right)$ is an
integer, the two approaches are exactly the same.

\iffullversion%
  \newcommand{\optnoisy}{Optimal partition selection as noisy thresholding}
\else%
  \newcommand{\optnoisy}{Noisy thresholding is optimal}
\fi%

\begin{theorem}[\optnoisy]\label{thm:noisy-thresholding}
  If $\del\in(0,1)$ and $\eps>0$ are such that
  $k=\frac{1}{\eps}\ln\left(\frac{\eeps+2\del-1}{(\eeps+1)\del}\right)$ is an
  integer, then for all $n$:
  \[
    \pop{n}=\proba{n+X \geq k+1}
  \]
  where $X$ is a random variable sampled from a $k$-truncated symmetric
  geometric distribution of success probability $(1-\emeps)$.
  \begin{proof}
    When $k = \frac{1}{\eps}\ln\left(\frac{\eeps+2\del-1}{(\eeps+1)\del}\right)$
    is an integer, we have $\ncr = k+1$, and
    \iffullversion%
      \begin{alignat*}{3}
        & & e^{k\eps} &= \frac{\eeps+2\del-1}{(\eeps+1)\del}\\
        &\implies &(e^{k\eps})(\eeps+1)\del &= \eeps-1 + 2\del\\
        &\implies &(e^{(k+1)\eps}-1)\delta + (e^{k\eps}-1)\del &= \eeps-1.
      \end{alignat*}
    \else%
      \[
        e^{k\eps} = \frac{\eeps+2\del-1}{(\eeps+1)\del}
      \]
      which leads to
      \[
        (e^{k\eps})(\eeps+1)\del = \eeps-1 + 2\del
      \]
      and
      \[
        (e^{(k+1)\eps}-1)\delta + (e^{k\eps}-1)\del = \eeps-1.
      \]
    \fi%

    On further rearranging, we get
    \[
      \frac{e^{(k+1)\eps}-1}{\eeps-1}\cdot\del + \frac{e^{k\eps}-1}{\eeps-1}\cdot\del = 1,
    \]
    and thus:
    \[
      1-\pop{\ncr} = \pop{\ncr-1}.
    \]

    From Lemma~\ref{lem:crossover-points}, we also get
    \[
      1-\pop{n} = \emeps((1-\pop{n-1})-\del)
    \]
    if $\ncr<n\leq\nf$. Since we also have
    \[
      \pop{n} = \eeps(\pop{n-1}+\del)
    \]
    if $0<n\leq\ncr$, we find that for $\ncr<n\leq\nf$,
    \begin{align*}
      \pop{n}
        & = 1 - \pop{2\ncr-1 -n} \\
        & = 1 - \frac{e^{(2k+1-n)\eps}-1}{\eeps-1}\cdot\del.
    \end{align*}
    Consequently, for such special combinations of $\eps$ and $\del$
    \[
      \nf= 2\ncr -1 = 2k+1.
    \]

    Now, rewriting the formula for $\popt$ in Theorem~\ref{thm:general-solution}
    using $\mu = n$ and $k = \ncr-1$ gives us
    \iffullversion%
      \[
        \pop{\mu} =
          \begin{cases}
            0
              & \textnormal{if } \mu \le 0 \\
            \frac{e^{(\mu+k+1-(k+1))\eps}-1}{\eeps-1}\cdot\del
              & \textnormal{if } \mu \le k+1 \\
            \left(1-e^{(\mu-(k+1))\eps}\right)\left(1+\frac{\del}{\eeps-1}\right)+e^{(\mu-(k+1))\eps}\frac{e^{(k+1)\eps}-1}{\eeps-1}\cdot\del
              & \textnormal{if } k+1 < n \le 2k+1 \\
            1 & \textnormal{otherwise.}
          \end{cases}
      \]
    \else%
      that $\pop{\mu}$ is:
      \begin{itemize}
        \item $0$ if $\mu\le0$,
        \item $\frac{e^{(\mu+k+1-(k+1))\eps}-1}{\eeps-1}\cdot\del$ if
          $\mu\le k+1$,
        \item $\left(1-e^{(\mu-(k+1))\eps}\right)\left(1+\frac{\del}{\eeps-1}\right)+e^{(\mu-(k+1))\eps}\frac{e^{(k+1)\eps}-1}{\eeps-1}\cdot\del$
          if $k+1<\mu\le2k+1$,
        \item $1$ otherwise.
      \end{itemize}
    \fi%

    Comparing this with~\eqref{eq:thresholding-prob} shows that for this
    combination of $\eps$ and $\del$, and for the corresponding derived values of
    $p$ and $k$,
    \[
      \pop{\mu} = \proba{Y \geq k+1}.
    \]
  \end{proof}
\end{theorem}

This characterization suggests a simple implementation of the optimal partition
selection primitive, at a minor cost in $\eps$ or $\del$. Given arbitrary $\eps$
and $\del$, one can replace $\eps$ by $\hat\eps\le\eps$, or $\del$ by
$\hat\del\le\del$ to ensure that $k$ from Theorem~\ref{thm:noisy-thresholding}
is an integer. In our definition of the truncated geometric mechanism, we choose
the latter strategy, requiring a slightly lower $\del$ by using an integer upper
bound on $k$, and using $p=1-e^{-\eps}$ to fully utilize the $\eps$ budget. We
then apply the truncated geometric mechanism to the number of unique users in
each partition, and return this partition if the noisy count is larger than $k$.
Further, this noisy count may also be published for such a partition, while
still satisfying $\epsdel$-differential privacy.

To see this, consider an arbitrarily large finite family of
partitions\footnote{For example, all possible partitions representable by
bytestrings that fit within available data storage} $Q$ such that each user in a
database $D$ is associated with at most one partition $q \in Q$. Consider the
following mechanism.

\begin{definition}[$k$-TSGD thresholded release]
For a database $D$, let $c_q(D)$ be the number of users associated with
partition $q$. Let $Q_D \subset Q$
be the finite subset $\setc{q}{q \in Q \text{ and } c_q(D) > 0}$
of partitions present in the dataset $D$.
Let the noise $X_q$ for $q \in Q$ be i.i.d.\ random variables drawn from
the $k$-TSGD  of parameters $p = 1-\emeps$ and
  $k = \left\lceil \frac{1}{\eps}\ln\left(\frac{\eeps+2\del-1}{(\eeps+1)\del}
  \right) \right\rceil$. Let $\hat{c}_q(D) = c_q(D) + X_q$.
Then, the $k$-TSGD thresholded release mechanism produces the set
\[
    \setc{(q,\hat{c}_q(D) )}{q \in Q_D \text{ and } \hat{c}_q(D) > k}.
\]
\end{definition}

\begin{theorem}
  The $k$-TSGD thresholded release mechanism satisfies $\epsdel$-differential
  privacy.
  \begin{proof}
    Consider the mechanism that adds a $k$-TSGD of parameter $p=1-\emeps$ and
    $k=\left\lceil\frac{1}{\eps}\ln\left(\frac{\eeps+2\del-1}{(\eeps+1)\del}\right)\right\rceil$
    to every possible partition count, including those not present in the
    dataset. That is, we apply the truncated geometric mechanism to the
    unique-user counts for all possible partitions (even partitions not
    contained in the database), which produces the set
    \[
      \setc{(q, \hat{c}_q(D))}{q \in Q}.
    \]

    This mechanism is $\epsdel$-differentially private: a single user's addition
    or removal changes only one partition, and on this partition,
    Theorem~\ref{thm:tgm-is-dp} shows that the output satisfies
    $\epsdel$-differential privacy. Combined with the condition that the noise
    values are independent, this means that the entire mechanism is also
    $\epsdel$-differentially private.

    Adding a thresholding step to release the noised values only when they are
    greater than $k$ is only post-processing. Therefore, the entire mechanism
    that releases
    \[
      \setc{(q, \hat{c}_q(D))}{ q \in Q \text{ and } \hat{c}_q(D) > k}
    \]
    is also $\epsdel$-differentially private.

    Now, notice that this mechanism is \emph{exactly the same} as if we had only
    added noise to the partitions in $Q_D$: the noise added to zero in empty
    partitions will be at most $k$, so these partitions will be removed from the
    output in the thresholding step. Since these two mechanisms are identical
    and one is $\epsdel$-differentially private, both are
    $\epsdel$-differentially private.
  \end{proof}
\end{theorem}

We note that this can be extended to the case where the set of allowed
partitions $Q$ is countably infinite, using standard techniques from measure
theory~\cite{holohan2015differential}. Thus, this mechanism is the
$\epsdel$-differential privacy equivalent of Algorithm \textsc{Filter}
in~\cite{cormode2012differentially}, which achieves $(\eps,0)$-differential
privacy when the set of possible partitions is known beforehand and not very
large.

To demonstrate the utility of such an operation, consider a slight variation of
the example query presented in the introduction.

\begin{sqllisting}
SELECT
  device_model,
  COUNT(user_id),
  AVG(latency)
FROM database
GROUP BY device_model
\end{sqllisting}

For simplicity, let us assume that each user contributes only one row with a
single value for the latency. Then, this may be implemented in the following
way.

\begin{minipage}{\linewidth}
\begin{sqllisting}
SELECT
  device_model,
  COUNT(user_id),
  SUM(latency) / COUNT(user_id)
FROM database
GROUP BY device_model
\end{sqllisting}
\end{minipage}

The available $\epsdel$ budget must be split up for the partition selection, sum
and count. Instead of instantiating separate noise values for partition
selection and the count and having to split up the $\epsdel$, we can use noisy
thresholding on the count. This may be used to obtain a more accurate count or
to leave more of the $\epsdel$ budget for the sum estimation.

\section{Numerical evaluation}

Theorem~\ref{thm:general-solution} shows that the optimal partition selection
primitive $\popt$ outperforms all other options. How does it compare with the
naive strategy of adding Laplace noise and thresholding the result?

\begin{definition}[Laplace partition selection~\cite{korolova2009releasing}]
  We denote by $\lap(b)$ a random variable sampled from a Laplace distribution
  of mean $0$ and of scale $b$. The following partition selection strategy
  $\slap$, called \emph{Laplace-based partition selection}, is
  $\epsdel$-differentially private:
  \[
  \slap(n) =
    \begin{cases}
      \drop & \textnormal{if } n+\lap\left(\frac{1}{\eps}\right) < 1-\frac{\ln(2\del)}{\eps} \\
      \keep & \textnormal{otherwise.}
    \end{cases}
  \]
  We denote by $\plap$ the corresponding partition selection primitive:
  $\plap(n)=\proba{\slap(n)=\keep}$.
\end{definition}

As expected, using the optimal partition selection primitive translates to a
larger probability of releasing a partition with the same user. As exemplified
in Figure~\ref{fig:proba-comparison}, the difference is especially large in the
high-privacy regime.

\iffullversion%
  \newcommand{\tikzgraph}[1]{\makebox[\textwidth][c]{#1}}
  \newcommand{\scalefigure}{1}
\else%
  \newcommand{\tikzgraph}[1]{\centering #1}
  \newcommand{\scalefigure}{0.9}
\fi%

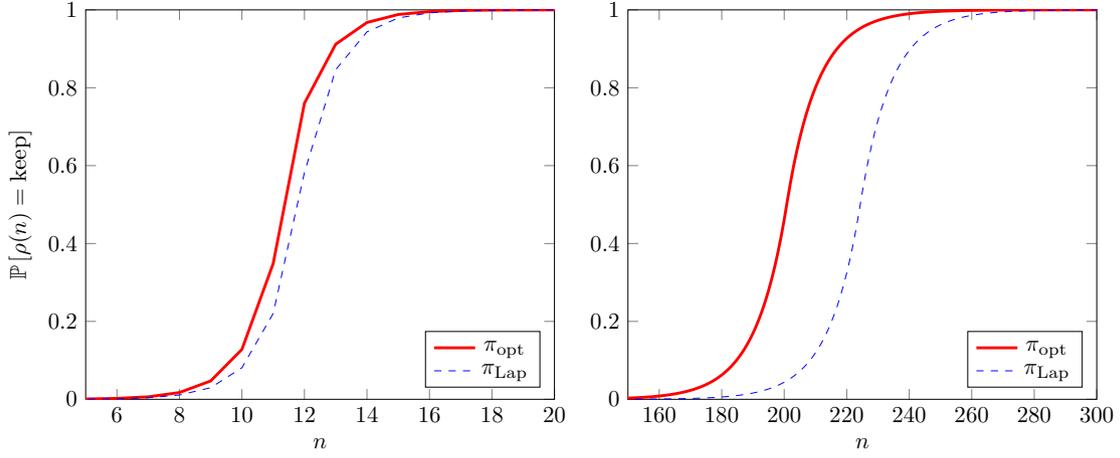
\begin{figure*}
  \tikzgraph{
    \begin{tikzpicture}[scale=\scalefigure]
      \begin{axis}[
          xmin=5,
          xmax=20,
          ymin=0,
          ymax=1,
          xlabel=$n$,
          ylabel=$\proba{\rho(n)=\keep}$,
          legend pos=south east,
        ]
        \addplot[
          red,
          very thick,
          solid
        ] table [x=n, y=opt, col sep=comma]{data/proba.eps1.del-5.csv};
        \addplot[
          blue,
          dashed
        ] table [x=n, y=lap, col sep=comma]{data/proba.eps1.del-5.csv};
        \legend{$\popt$,$\plap$}
      \end{axis}
    \end{tikzpicture}
    \begin{tikzpicture}[scale=\scalefigure]
      \begin{axis}[
          xmin=150,
          xmax=300,
          ymin=0,
          ymax=1,
          xlabel=$n$,
          legend pos=south east,
        ]
        \addplot[
          red,
          very thick,
          solid
        ] table [x=n, y=opt, col sep=comma]{data/proba.eps0.1.del-10.csv};
        \addplot[
          blue,
          dashed
        ] table [x=n, y=lap, col sep=comma]{data/proba.eps0.1.del-10.csv};
        \legend{$\popt$,$\plap$}
      \end{axis}
    \end{tikzpicture}
  }
  \caption{Probability of releasing a partition depending on the number of
  unique users, comparing Laplace-based partition selection with $\popt$. On the
  left, $\eps=1$ and $\del=10^{-5}$, on the right, $\eps=0.1$ and
  $\del=10^{-10}$.}%
  \label{fig:proba-comparison}
\end{figure*}

\begin{figure*}
  \tikzgraph{
    \begin{tikzpicture}[scale=\scalefigure]
      \begin{axis}[
          xmode=log,
          xmin=1e-12,
          xmax=1e-3,
          xtick={1e-12, 1e-10, 1e-8, 1e-6, 1e-4},
          try min ticks=5,
          ymin=0,
          ymax=300,
          xlabel=$\del$,
          ylabel={$n$ s.t. $\proba{\rho(n)=\keep}=0.05,0.50,0.95$},
          no markers
        ]
        \addplot[
          red,
          very thick,
          dashed
        ] table [x=del, y=opt05, col sep=comma]{data/midpoint.eps0.1.revisited.csv};
        \addplot[
          red,
          very thick,
          solid
        ] table [x=del, y=opt50, col sep=comma]{data/midpoint.eps0.1.revisited.csv};
        \addplot[
          red,
          very thick,
          dashed
        ] table [x=del, y=opt95, col sep=comma]{data/midpoint.eps0.1.revisited.csv};
        \addplot[
          blue,
          dashed
        ] table [x=del, y=lap05, col sep=comma]{data/midpoint.eps0.1.revisited.csv};
        \addplot[
          blue,
          solid
        ] table [x=del, y=lap50, col sep=comma]{data/midpoint.eps0.1.revisited.csv};
        \addplot[
          blue,
          dashed
        ] table [x=del, y=lap95, col sep=comma]{data/midpoint.eps0.1.revisited.csv};
        \legend{, $\popt$, , , $\plap$,}
      \end{axis}
    \end{tikzpicture}
    \begin{tikzpicture}[scale=\scalefigure]
      \begin{axis}[
          xmode=log,
          xmin=0.000000000001,
          xmax=0.001,
          xtick={1e-12, 1e-10, 1e-8, 1e-6, 1e-4},
          ymin=0,
          ymax=30,
          xlabel=$\del$,
          no markers
        ]
        \addplot[
          red,
          very thick,
          dashed
        ] table [x=del, y=opt05, col sep=comma]{data/midpoint.eps1.revisited.csv};
        \addplot[
          red,
          very thick,
          solid
        ] table [x=del, y=opt50, col sep=comma]{data/midpoint.eps1.revisited.csv};
        \addplot[
          red,
          very thick,
          dashed
        ] table [x=del, y=opt95, col sep=comma]{data/midpoint.eps1.revisited.csv};
        \addplot[
          blue,
          dashed
        ] table [x=del, y=lap05, col sep=comma]{data/midpoint.eps1.revisited.csv};
        \addplot[
          blue,
          solid
        ] table [x=del, y=lap50, col sep=comma]{data/midpoint.eps1.revisited.csv};
        \addplot[
          blue,
          dashed
        ] table [x=del, y=lap95, col sep=comma]{data/midpoint.eps1.revisited.csv};
        \legend{, $\popt$, , , $\plap$,}
      \end{axis}
    \end{tikzpicture}
  }
  \caption{Comparison of the 5th, 50th and 95th percentile of the partition
    strategy $\rho$ as a function of $\del$, for $\eps=0.1$ (left) and $\eps=1$
    (right). The mid-point is plotted as a solid line, while the 5th and 95th
    percentiles are dashed.}%
  \label{fig:midpoint-delta}
\end{figure*}
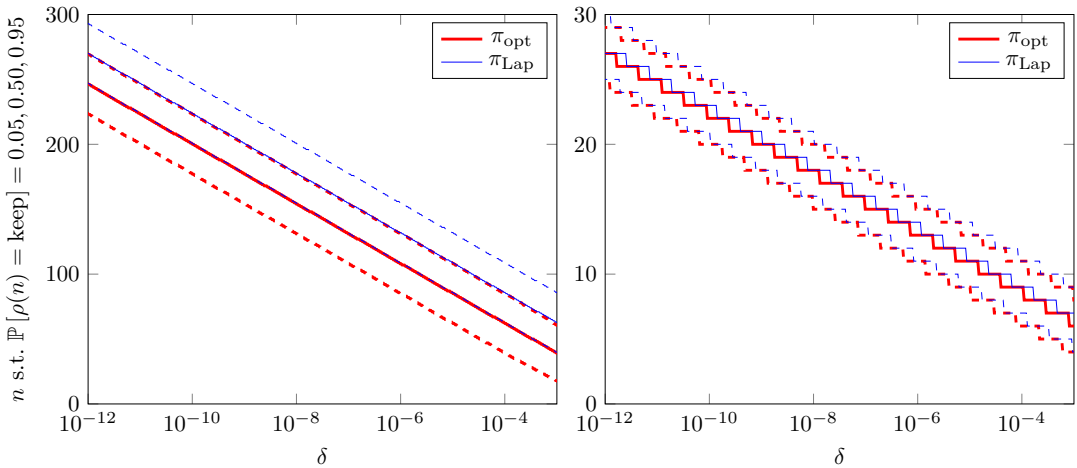

To better understand the dependency on $\eps$ and $\del$, we also compare the
\emph{midpoint} obtained for both partition selection strategies $\rho$: the
number $n$ for which the probability of releasing a partition with $n$ users is
$0.5$. For Laplace-based partition selection, this $n$ is simply the threshold.
As Figure~\ref{fig:midpoint-epsilon} shows, the gains are especially substantial
when $\eps$ is small, and not significant for $\eps>1$.
Figure~\ref{fig:midpoint-delta} shows the dependency on $\del$: for a fixed
$\eps$, there is a \emph{constant} interval between the midpoints of both
strategies. Thus, the relative gains are larger for a larger $\del$, since the
midpoint is also smaller.

\begin{figure*}
  \tikzgraph{
    \begin{tikzpicture}[scale=\scalefigure]
      \begin{axis}[
          xmode=log,
          xmin=0.01,
          xmax=0.5,
          log ticks with fixed point,
          ymin=0,
          ymax=1000,
          xlabel=$\eps$,
          ylabel={$n$ s.t. $\proba{\rho(n)=\keep}=0.05,0.50,0.95$},
          no markers,
        ]
        \addplot[
          red,
          very thick,
          dashed
        ] table [x=eps, y=opt05, col sep=comma]{data/midpoint.del-5.revisited.csv};
        \addplot[
          red,
          very thick,
          solid
        ] table [x=eps, y=opt50, col sep=comma]{data/midpoint.del-5.revisited.csv};
        \addplot[
          red,
          very thick,
          dashed
        ] table [x=eps, y=opt95, col sep=comma]{data/midpoint.del-5.revisited.csv};
        \addplot[
          blue,
          dashed
        ] table [x=eps, y=lap05, col sep=comma]{data/midpoint.del-5.revisited.csv};
        \addplot[
          blue,
          solid
        ] table [x=eps, y=lap50, col sep=comma]{data/midpoint.del-5.revisited.csv};
        \addplot[
          blue,
          dashed
        ] table [x=eps, y=lap95, col sep=comma]{data/midpoint.del-5.revisited.csv};
        \legend{, $\popt$, , , $\plap$,}
      \end{axis}
    \end{tikzpicture}
    \begin{tikzpicture}[scale=\scalefigure]
      \begin{axis}[
          xmode=log,
          xmin=0.1,
          xmax=3,
          log ticks with fixed point,
          ymin=0,
          ymax=100,
          xlabel=$\eps$,
          no markers
        ]
        \addplot[
          red,
          very thick,
          dashed
        ] table [x=eps, y=opt05, col sep=comma]{data/midpoint.del-5.revisited.csv};
        \addplot[
          red,
          very thick,
          solid
        ] table [x=eps, y=opt50, col sep=comma]{data/midpoint.del-5.revisited.csv};
        \addplot[
          red,
          very thick,
          dashed
        ] table [x=eps, y=opt95, col sep=comma]{data/midpoint.del-5.revisited.csv};
        \addplot[
          blue,
          dashed
        ] table [x=eps, y=lap05, col sep=comma]{data/midpoint.del-5.revisited.csv};
        \addplot[
          blue,
          solid
        ] table [x=eps, y=lap50, col sep=comma]{data/midpoint.del-5.revisited.csv};
        \addplot[
          blue,
          dashed
        ] table [x=eps, y=lap95, col sep=comma]{data/midpoint.del-5.revisited.csv};
        \legend{, $\popt$, , , $\plap$,}
      \end{axis}
    \end{tikzpicture}
  }
  \caption{Comparison of the 5th, 50th and 95th percentile of the partition
    selection strategy $\rho$ as a function of $\eps$, for $\del=10^{-5}$. The
    mid-point is plotted as a solid line, while the 5th and 95th percentiles are
    dashed.}%
  \label{fig:midpoint-epsilon}
\end{figure*}
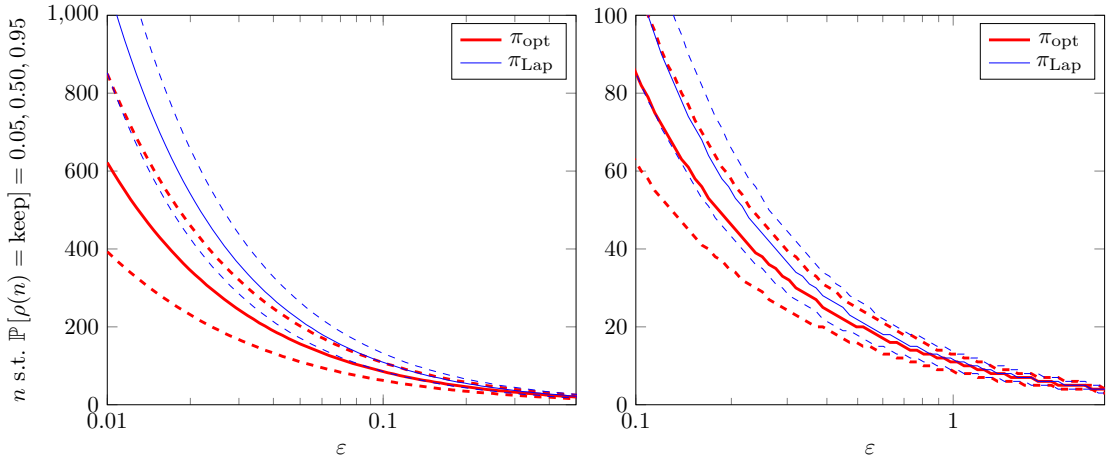

We also verified experimentally that on each partition, the selection mechanism
runs in constant, very short time, on the order of 100 nanoseconds on a standard
machine. This is not surprising: Theorem~\ref{thm:general-solution} provides a
simple, closed-form formula for computing $\popt(n)$, and generating the random
decision based on this probability is computationally trivial. The performance
impact of Laplace-based thresholding is similarly negligible: the only real cost
of such simple partition selection strategies is to count the number of unique
users $n$ in each partition, which is orders of magnitude more computationally
intensive than computing $\pi(n)$.

\section{Discussion}

The approach presented in this work is both easy to implement and efficient.
Counting the number of unique users per partition can be done in one pass over
the data and is massively parallelizable. Furthermore, since there is a
relatively small value $k$ such that the probability of keeping a partition with
$n \ge k$ users is 1, the counting process can be interrupted as soon as a
partition reaches $k$ users. This keeps memory usage low (in $O(k)$) without
requiring approximate count-distinct algorithms like
HyperLogLog~\cite{flajolet2007hyperloglog} for which a more complex sensitivity
analysis would be needed.

\paragraph*{Extension to multiple partitions per user}

Our approach could, in principle, be extended to cases where each user can
contribute to $\kappa>1$ partitions. Following the intuition of
Lemma~\ref{lem:recursive-solution}, we could list a set of recursive equations
defining $\pop{n}$ as a function of $\popt(i)$ for $i<n$. Sadly, the system of
equations quickly gets too large to solve naively. Consider, for example, the
case where $\kappa=2$. The differential privacy constraint requires, for all
$n \ge i \ge 0$ and all $S\subseteq\dk^2$:
\begin{align*}
  \iffullversion%
    \proba{\left(\strat(n+1),\strat(i+1)\right)\in S}
      & \leq \eeps\cdot\proba{\left(\strat(n),\strat(i)\right)\in S} + \del \\
    \proba{\left(\strat(n),\strat(i)\right)\in S}
      & \leq \eeps\cdot\proba{\left(\strat(n+1),\strat(i+1)\right)\in S} + \del
  \else%
    & \proba{\left(\strat(n+1),\strat(i+1)\right)\in S} \\
    & \qquad \leq \eeps\cdot\proba{\left(\strat(n),\strat(i)\right)\in S} + \del \\
    & \proba{\left(\strat(n),\strat(i)\right)\in S} \\
    & \qquad \leq \eeps\cdot\proba{\left(\strat(n+1),\strat(i+1)\right)\in S} + \del
  \fi%
\end{align*}

Thus, to maximize $\pi(n)$, we have to consider $32n$ inequalities: $n$ possible
values of $i$, $2^{\left(2^2\right)}=16$ possible values of $S$, and two
inequalities. When $\kappa$ increases, the total number of inequalities to
compute all elements up to $n$ is $O\left(n^{\kappa}2^{\kappa^2}\right)$. Some
of these inequalities are trivial (e.g., when $S=\emptyset$ or $S=\dk^\kappa$),
but most are not. We do not know whether it is possible to only consider a small
number of these inequalities, and obtain the others ``for free''.

Furthermore, the recurrence-based proof of optimality of $\popt$ only holds when
we assume that each user contributes to \emph{exactly} $\kappa$ partitions in
the original dataset. As discussed previously, this is relatively frequent when
$\kappa=1$, but it rarely happens for larger values of $\kappa$: in typical
datasets, some users contribute to more partitions than others. In that case,
\emph{weighing} the contributions of each user differently can bring additional
benefits, as can changing each user's strategy based on those of previous
users~\cite{gopi2020differentially}. For this generalized problem, it seems
difficult to even define what optimality means.

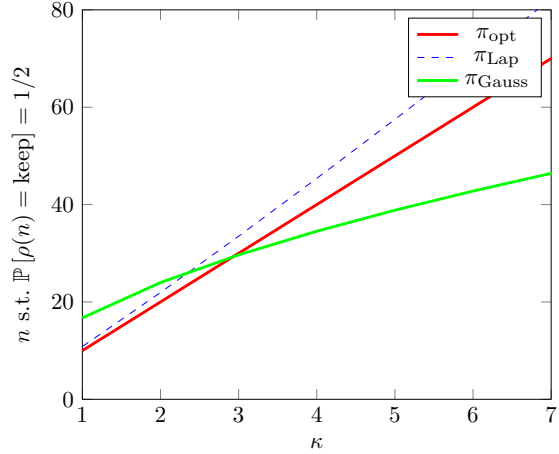
\begin{figure*}
  \tikzgraph{
    \begin{tikzpicture}[scale=\scalefigure]
      \begin{axis}[
          xmin=1,
          xmax=7,
          ymin=0,
          ymax=80,
          xlabel=$\kappa$,
          ylabel={$n$ s.t. $\proba{\rho(n)=\keep}=1/2$},
          no markers
        ]
        \addplot[
          red,
          very thick,
          solid
        ] table [x=kappa, y=opt-naive, col sep=comma]{data/morecontributions.eps1.del-5.csv};
        \addplot[
          blue,
          dashed
        ] table [x=kappa, y=lap-naive, col sep=comma]{data/morecontributions.eps1.del-5.csv};
        \addplot[
          green,
          very thick,
          solid
        ] table [x=kappa, y=gauss-naive, col sep=comma]{data/morecontributions.eps1.del-5.csv};
        \legend{$\popt$,$\plap$,$\pgauss$}
      \end{axis}
    \end{tikzpicture}
  }
  \caption{Comparison of the mid-point of the partition selection strategy
    $\rho$ as a function of $\kappa$, for $\eps=1$ and $\delta=10^{-5}$. For
    $\kappa>1$, the privacy budget is divided by $\kappa$ for $\popt$ and
    $\plap$; for $\pgauss$, we use the formula in~\cite{balle2018improving} to
    derive the standard deviation of Gaussian noise, and we split the $\delta$
    between noise and thresholding to minimize the threshold.}%
  \label{fig:midpoint-kappa}
\end{figure*}

The simplest option to use our approach for $\kappa>1$ is to divide the total
privacy budget in $\kappa$. For generic tooling with strict scalability
requirements where the analyst manually specifies $\kappa$, we recommend using
our method (splitting the privacy budget) for $\kappa\le3$, and weighted
Gaussian thresholding (described in~\cite{gopi2020differentially}) for
$\kappa\ge4$. Figure~\ref{fig:midpoint-kappa} compares the mid-point of the
partition selection strategy between $\popt$, Laplace-based thresholding, and
(non-weighted) Gaussian-based thresholding. It shows that the crossing point
happens for $\kappa=3$, this stays true for varying values $\eps$ and $\delta$.

Comparison with weighted Gaussian thresholding is less trivial, since its
benefits depend on the data distribution. However, weighted Gaussian
thresholding is always better than Gaussian-based thresholding, and is
straightforward to implement in a massively parallelizable fashion. We have also
observed that its utility benefits are only significant for large $\kappa$, so
our recommendation to use $\popt$ for $\kappa\le3$ is likely robust.

Policy-based approaches like those described in~\cite{gopi2020differentially}
also provide more utility, but they are not as scalable: since the strategy for
each user depends on the choices made by all previous users, the computation
cannot be parallelized. It also requires to keep an in-memory histogram of all
partitions seen so far, which also does not scale to extremely large datasets.
Improving the scalability of such policy-based approaches is an interesting open
problem, on which further research would be valuable.

\paragraph*{Extension to bounded differential privacy}

In the definition of differential privacy we use in this work, neighboring
databases differ in a single user being added or removed. This notion is called
\emph{unbounded} differential privacy in~\cite{kifer2011no}, by contrast to
\emph{bounded} differential privacy, in which neighboring datasets differ in a
single user \emph{changing} their data. $\epsdel$-unbounded DP implies
$(2\eps,2\del)$-DP, which provides a trivial way to extend our method to the
bounded version of the definition: simply divide the privacy budget by two.
This method outperforms Laplace-based thresholding, since Laplace noise of scale
$2/\eps$ must be added in the bounded setting (since $L_1$-sensitivity is $2$
and no longer $1$). Further, when $k$ from Theorem~\ref{thm:noisy-thresholding}
is an integer, this noise distribution exactly achieves the lower bound on the
loss from~\cite{geng2016adp}, and is therefore optimal for arbitrary symmetric
loss functions.

\paragraph*{Extension to weighted sampling}

Another extension to this work was developed independently
in~\cite{cgss21}: the authors consider the problem of \emph{differentially
private weighted sampling}, where the partitions to be selected are also sampled
to generate a compact summary of the data, rather than the maximal dataset. In
the extreme case when the sampling probability is one, this is equivalent to
partition selection.

\paragraph*{Other possible extensions}

The truncated geometric mechanism can be used as a building block to replace
the Laplace or geometric mechanism in situations where $\epsdel$-DP with
$\del>0$ is acceptable. Similarly to the truncated Laplace
mechanism~\cite{geng2020tight}, this building block is optimal for
integer-valued functions.

To see how such a building block could be used in practice, consider the problem
of releasing a histogram where some partitions are known in advance (call them
\emph{public partitions}), and some are not and must be discovered using the
private data (\emph{private partitions}). Note that some public partitions might
be absent from the private data. In that case, one could add truncated geometric
noise to all partitions (public \emph{and} private), and use two distinct
thresholds: one given by the formula for $k$ in Definition~\ref{def:tgm}, and an
arbitrary one $t$.
\begin{itemize}
  \item $k$ is used to threshold the partitions present in the private data
    \emph{but not in the list of public partitions};
  \item $t$ is used to threshold the public partitions (whether or not they are
    also in the private data).
\end{itemize}
The second threshold $t$ can be arbitrary, and allows an analyst to control the
trade-off between false positives and false negatives. For example, setting
$t=0$ guarantees that all public partitions that appear in the private data are
present in the output (no false negatives), at the cost of having a potentially
large number of public partitions appearing in the output even though they were
not present (many false positives). Conversely, setting $t=k$ guarantees that
\emph{only} the partitions present in the private data can be present in the
output (no false positives), at the cost of dropping potentially many of them
(many false negatives). Intermediate values of $t$ can allow an analyst to more
finely tune this trade-off depending on the application.

We postulate that this building block could be used in a variety of different
settings, and combined with existing techniques. For example, one could build a
variant of the standard vector technique~\cite{dwork2009complexity} that uses
the truncated geometric mechanism instead of the Laplace mechanism to add noise
to the output of the queries and to the threshold. This could be used to
efficiently simulate the standard vector technique on a very large number of
queries, most of which are deterministically below the threshold and can be
skipped during computation. Formalizing this intuition and using it for
partition selection with $\kappa>1$ is left to future work.

\section{Conclusion}

We introduced an optimal primitive for differentially private partition
selection, a special case of differentially private set union where the
sensitivity is $1$. This optimal approach is simple to implement and efficient.
It outperforms Laplace-based thresholding; the utility gain is especially
significant in the high-privacy (small $\eps$) regime. Besides the possible
research directions outlined previously, this work leaves two open questions. Is
it possible to extend this optimal approach to larger sensitivities in a simple
and efficient manner? Furthermore, is it possible to combine this primitive with
existing approaches to differentially private set
union~\cite{gopi2020differentially}, like weighted histograms or policy-based
strategies?

\section{Acknowledgments}
The authors gratefully acknowledge Alex Kulesza, Chao Li, Michael Daub, Kareem
Amin, Peter Dickman, Peter Kairouz, and the PETS reviewers for their helpful
feedback on this work.

D.D.\ was employed by Google and ETH Zurich at the time of this work. This
research received no specific grant from any funding agency in the public,
commercial, or not-for-profit sector.

\bibliographystyle{alpha}
\bibliography{biblio}

\newcommand{\etalchar}[1]{$^{#1}$}
\begin{thebibliography}{KKMN09}

\bibitem[ACC12]{acs2012differentially}
Gergely Acs, Claude Castelluccia, and Rui Chen.
\newblock Differentially private histogram publishing through lossy
  compression.
\newblock In {\em 2012 IEEE 12th International Conference on Data Mining},
  pages 1--10. IEEE, 2012.

\bibitem[BHE{\etalchar{+}}18]{bater2018shrinkwrap}
Johes Bater, Xi~He, William Ehrich, Ashwin Machanavajjhala, and Jennie Rogers.
\newblock Shrinkwrap: Differentially-private query processing in private data
  federations.
\newblock {\em arXiv preprint arXiv:1810.01816}, 2018.

\bibitem[BW18]{balle2018improving}
Borja Balle and Yu-Xiang Wang.
\newblock Improving the gaussian mechanism for differential privacy: Analytical
  calibration and optimal denoising.
\newblock In {\em International Conference on Machine Learning}, pages
  394--403. PMLR, 2018.

\bibitem[CGSS21]{cgss21}
Edith Cohen, Ofir Geri, Tamas Sarlos, and Uri Stemmer.
\newblock Differentially private weighted sampling.
\newblock In Arindam Banerjee and Kenji Fukumizu, editors, {\em Proceedings of
  The 24th International Conference on Artificial Intelligence and Statistics},
  volume 130 of {\em Proceedings of Machine Learning Research}, pages
  2404--2412. PMLR, 13--15 Apr 2021.

\bibitem[CPST11]{cormode2011differentially}
Graham Cormode, Magda Procopiuc, Divesh Srivastava, and Thanh~TL Tran.
\newblock Differentially private publication of sparse data.
\newblock {\em arXiv preprint arXiv:1103.0825}, 2011.

\bibitem[CPST12]{cormode2012differentially}
Graham Cormode, Cecilia Procopiuc, Divesh Srivastava, and Thanh~TL Tran.
\newblock Differentially private summaries for sparse data.
\newblock In {\em Proceedings of the 15th International Conference on Database
  Theory}, pages 299--311, 2012.

\bibitem[DMNS06]{dwork2006calibrating}
Cynthia Dwork, Frank McSherry, Kobbi Nissim, and Adam Smith.
\newblock Calibrating noise to sensitivity in private data analysis.
\newblock In {\em Theory of Cryptography Conference}, pages 265--284. Springer,
  2006.

\bibitem[DNR{\etalchar{+}}09]{dwork2009complexity}
Cynthia Dwork, Moni Naor, Omer Reingold, Guy~N Rothblum, and Salil Vadhan.
\newblock On the complexity of differentially private data release: efficient
  algorithms and hardness results.
\newblock In {\em Proceedings of the forty-first annual ACM symposium on Theory
  of computing}, pages 381--390, 2009.

\bibitem[DR14]{dwork2014algorithmic}
Cynthia Dwork and Aaron Roth.
\newblock The algorithmic foundations of differential privacy.
\newblock {\em Foundations and Trends in Theoretical Computer Science},
  9(3-4):211--407, 2014.

\bibitem[DWHL11]{ding2011differentially}
Bolin Ding, Marianne Winslett, Jiawei Han, and Zhenhui Li.
\newblock Differentially private data cubes: optimizing noise sources and
  consistency.
\newblock In {\em Proceedings of the 2011 ACM SIGMOD International Conference
  on Management of data}, pages 217--228, 2011.

\bibitem[FFGM07]{flajolet2007hyperloglog}
Philippe Flajolet, {\'E}ric Fusy, Olivier Gandouet, and Fr{\'e}d{\'e}ric
  Meunier.
\newblock Hyperloglog: the analysis of a near-optimal cardinality estimation
  algorithm.
\newblock In {\em Discrete Mathematics and Theoretical Computer Science}, pages
  137--156. Discrete Mathematics and Theoretical Computer Science, 2007.

\bibitem[GDGK20]{geng2020tight}
Quan Geng, Wei Ding, Ruiqi Guo, and Sanjiv Kumar.
\newblock Tight analysis of privacy and utility tradeoff in approximate
  differential privacy.
\newblock In {\em International Conference on Artificial Intelligence and
  Statistics}, pages 89--99, 2020.

\bibitem[GGK{\etalchar{+}}20]{gopi2020differentially}
Sivakanth Gopi, Pankaj Gulhane, Janardhan Kulkarni, Judy~Hanwen Shen, Milad
  Shokouhi, and Sergey Yekhanin.
\newblock Differentially private set union.
\newblock {\em arXiv preprint arXiv:2002.09745}, 2020.

\bibitem[GRS12]{ghosh2012universally}
Arpita Ghosh, Tim Roughgarden, and Mukund Sundararajan.
\newblock Universally utility-maximizing privacy mechanisms.
\newblock {\em SIAM Journal on Computing}, 41(6):1673--1693, 2012.

\bibitem[GV16]{geng2016adp}
Quan Geng and Pramod Viswanath.
\newblock Optimal noise adding mechanisms for approximate differential privacy.
\newblock {\em IEEE Transactions on Information Theory}, 62(2):952--969, Feb
  2016.

\bibitem[HLM15]{holohan2015differential}
Naoise Holohan, Douglas~J Leith, and Oliver Mason.
\newblock Differential privacy in metric spaces: Numerical, categorical and
  functional data under the one roof.
\newblock {\em Information Sciences}, 305:256--268, 2015.

\bibitem[HRMS09]{hay2009boosting}
Michael Hay, Vibhor Rastogi, Gerome Miklau, and Dan Suciu.
\newblock Boosting the accuracy of differentially-private histograms through
  consistency.
\newblock {\em arXiv preprint arXiv:0904.0942}, 2009.

\bibitem[IK06]{inusah2006discrete}
Seidu Inusah and Tomasz~J Kozubowski.
\newblock A discrete analogue of the laplace distribution.
\newblock {\em Journal of statistical planning and inference},
  136(3):1090--1102, 2006.

\bibitem[JNS18]{johnson2018towards}
Noah Johnson, Joseph~P Near, and Dawn Song.
\newblock Towards practical differential privacy for sql queries.
\newblock {\em Proceedings of the VLDB Endowment}, 11(5):526--539, 2018.

\bibitem[KKMN09]{korolova2009releasing}
Aleksandra Korolova, Krishnaram Kenthapadi, Nina Mishra, and Alexandros
  Ntoulas.
\newblock Releasing search queries and clicks privately.
\newblock In {\em Proceedings of the 18th international conference on World
  wide web}, pages 171--180, 2009.

\bibitem[KM11]{kifer2011no}
Daniel Kifer and Ashwin Machanavajjhala.
\newblock No free lunch in data privacy.
\newblock In {\em Proceedings of the 2011 ACM SIGMOD International Conference
  on Management of data}, pages 193--204, 2011.

\bibitem[KMS20]{kaplan2020sparse}
Haim Kaplan, Yishay Mansour, and Uri Stemmer.
\newblock The sparse vector technique, revisited.
\newblock {\em arXiv preprint arXiv:2010.00917}, 2020.

\bibitem[KTH{\etalchar{+}}19]{kotsogiannis2019privatesql}
Ios Kotsogiannis, Yuchao Tao, Xi~He, Maryam Fanaeepour, Ashwin Machanavajjhala,
  Michael Hay, and Gerome Miklau.
\newblock Privatesql: a differentially private sql query engine.
\newblock {\em Proceedings of the VLDB Endowment}, 12(11):1371--1384, 2019.

\bibitem[LC14]{lee2014top}
Jaewoo Lee and Christopher~W Clifton.
\newblock Top-k frequent itemsets via differentially private fp-trees.
\newblock In {\em Proceedings of the 20th ACM SIGKDD international conference
  on Knowledge discovery and data mining}, pages 931--940, 2014.

\bibitem[LSL16]{lyu2016understanding}
Min Lyu, Dong Su, and Ninghui Li.
\newblock Understanding the sparse vector technique for differential privacy.
\newblock {\em arXiv preprint arXiv:1603.01699}, 2016.

\bibitem[LXJ14]{li2014differentially}
Haoran Li, Li~Xiong, and Xiaoqian Jiang.
\newblock Differentially private synthesization of multi-dimensional data using
  copula functions.
\newblock In {\em Advances in database technology: proceedings. International
  conference on extending database technology}, volume 2014, page 475. NIH
  Public Access, 2014.

\bibitem[WZL{\etalchar{+}}19]{wilson2019differentially}
Royce~J Wilson, Celia~Yuxin Zhang, William Lam, Damien Desfontaines, Daniel
  Simmons-Marengo, and Bryant Gipson.
\newblock Differentially private {SQL} with bounded user contribution.
\newblock {\em arXiv preprint arXiv:1909.01917}, 2019.

\bibitem[XWG10]{xiao2010differential}
Xiaokui Xiao, Guozhang Wang, and Johannes Gehrke.
\newblock Differential privacy via wavelet transforms.
\newblock {\em IEEE Transactions on knowledge and data engineering},
  23(8):1200--1214, 2010.

\bibitem[XXFG12]{xiao2012dpcube}
Yonghui Xiao, Li~Xiong, Liyue Fan, and Slawomir Goryczka.
\newblock Dpcube: differentially private histogram release through
  multidimensional partitioning.
\newblock {\em arXiv preprint arXiv:1202.5358}, 2012.

\bibitem[XZX{\etalchar{+}}13]{xu2013differentially}
Jia Xu, Zhenjie Zhang, Xiaokui Xiao, Yin Yang, Ge~Yu, and Marianne Winslett.
\newblock Differentially private histogram publication.
\newblock {\em The VLDB Journal}, 22(6):797--822, 2013.

\bibitem[ZCP{\etalchar{+}}17]{zhang2017privbayes}
Jun Zhang, Graham Cormode, Cecilia~M Procopiuc, Divesh Srivastava, and Xiaokui
  Xiao.
\newblock Privbayes: Private data release via bayesian networks.
\newblock {\em ACM Transactions on Database Systems (TODS)}, 42(4):1--41, 2017.

\end{thebibliography}

\end{document}